\documentclass{pasj00}
\draft

\begin{document}
\SetRunningHead{K. Fujisawa et al.}{Periodic Flare of the 6.7~GHz Methanol Maser in IRAS~22198$+$6336}
%\Received{}%{yyyy/mm/dd}
%\Accepted{}%{yyyy/mm/dd}
%\Published{}%{yyyy/mm/dd}

\title{Periodic Flare of the 6.7~GHz Methanol Maser in IRAS~22198$+$6336}

%%% begin:list of authors
% Do NOT capitalize all letters in "textsc".
\author{%
  Kenta \textsc{Fujisawa},\altaffilmark{1,2}
  Genta \textsc{Takase},\altaffilmark{1}
  Saki \textsc{Kimura},\altaffilmark{1}
  Nozomu \textsc{Aoki},\altaffilmark{1}
  Yoshito \textsc{Nagadomi},\altaffilmark{1}
  Tadashi \textsc{Shimomura},\altaffilmark{1}
  Koichiro \textsc{Sugiyama},\altaffilmark{1}
  Kazuhito \textsc{Motogi},\altaffilmark{1}
  Kotaro \textsc{Niinuma},\altaffilmark{1}
  Tomoya \textsc{Hirota},\altaffilmark{3}
  and
  Yoshinori \textsc{Yonekura}\altaffilmark{4}}
\altaffiltext{1}{Department of Physics, Faculty of Science, Yamaguchi University, Yoshida 1677-1, Yamaguchi-city, Yamaguchi 753-8512}
\altaffiltext{2}{The Research Institute of Time Studies, Yamaguchi University, Yoshida 1677-1, Yamaguchi-city, Yamaguchi 753-8511}
\altaffiltext{3}{Mizusawa VLBI Observatory, National Astronomical Observatory of Japan, Hoshigaoka-cho 2-12, Oshu, Iwate 023-0861}
\altaffiltext{4}{Center for Astronomy, Ibaraki University, 2-1-1 Bunkyo, Mito, Ibaraki 310-8512}
\email{kenta@yamaguchi-u.ac.jp}
%%% end:list of authors

%%% Please use the following style in case that sorting by 
%%% affiliation is impossible. 
%
% \author{%
%   D-Firstname \textsc{D-Familyname}\altaffilmark{1}
%   E-Firstname \textsc{E-Familyname}\altaffilmark{1,2}
%   and
%   F-Firstname \textsc{F-Familyname}\altaffilmark{2}}
% \altaffiltext{1}{Address of Institute}
% \email{ddddd@xxx.xxx.xx.xx}
% \email{eeeee@xxx.xxx.xx.xx}
% \altaffiltext{2}{Address of Institute}

%% `\KeyWords{}' always has to be placed before `\maketitle'.
\KeyWords{ISM: Star forming regions --- ISM: individual (IRAS~22198$+$6336) --- masers: methanol} %Do NOT move this preamble from here!

\maketitle

\begin{abstract}
We have detected periodic flares of the 6.7~GHz methanol maser from
an intermediate-mass star-forming region IRAS~22198$+$6336.
The maser was monitored daily in 2011, 2012, and 2013.
Six flares were observed with a period of 34.6~days.
The variation pattern is intermittent, and the flux ratio of
the flaring and the quiescent states exceeds 30.
Such intermittent variation with the short period uniquely
characterizes the variation of the IRAS~22198$+$6336 maser.
At least five spectral components were identified.
The spectral components varied almost synchronously,
but their peak times differed by 1.8~days.
These characteristics can be explained by the
colliding-wind binary model.
\end{abstract}

\section{Introduction}
The 6.7~GHz methanol maser is emitted from high-mass star-forming regions
(e.g., Menten 1991; Caswell et al. 1995; Minier et al. 2003;
Xu et al. 2008; Breen et al. 2013).
The maser is considered to trace the gas disk or outflow of
young stellar objects (e.g., Minier et al. 2000; De Buizer 2003;
Sugiyama et al. 2014). Since Goedhart et al. (2003, 2004)
first demonstrated flux variation in this maser, several
variability patterns, including periodicity, have been recognized.
The first detected periodic source was G9.62$+$0.20E, with a period
of 246~days. To date, periodicity has been reported for 11 sources,
with periods ranging from 29.5~days for G12.89$+$0.49 to 668~days for
G196.45$-$1.68 (Araya et al. 2010; Goedhart et al. 2003, 2004, 2007,
2009; Szymczak et al. 2011). Some of these sources
(e.g., G12.89$+$0.49) show continuous, sinusoidal variation,
while others (e.g., G37.55$+$0.20) display flaring variation
(with intermittently rising flux). Sugiyama et al. (2008)
reported flux variation in the aperiodic source Cep A and
studied the excitation mechanism and spatial distribution
of its 6.7~GHz methanol maser. Fujisawa et al. (2012)
discovered a bursting maser with a timescale of less than one
day in G33.64$-$0.21, and they proposed its source as release of
local magnetic energy in the gas disk. Thus, by investigating
short-term flux variations of the 6.7~GHz methanol maser,
we can better understand the circumstellar environments in
star-forming regions.

This paper reports our observations of the 6.7~GHz methanol maser in IRAS~22198$+$6336.
The systemic velocity of this source is $-11$~km~s$^{-1}$ (Tafalla et al. 1993)
which is close to the velocities of the maser components.
The source is located in a Cepheus$-$Cassiopeia molecular cloud
complex (Yonekura et al. 1997), and accompanies the dark nebula
Lynds 1204G (Sanchez-Monge et al. 2008) closely located to an H\emissiontype{II} region S140.
From annual parallax
observations, the distance of IRAS~22198$+$6336 was determined as
$764 \pm 27$ pc (Hirota et al. 2008).
Consequently, its luminosity and mass were derived as
$450 L_{\solar}$ and $7 M_{\solar}$ (Hirota et al. 2008) or
$370 L_{\solar}$ and $5 M_{\solar}$ (Sanchez-Monge et al. 2010), respectively.
Given its small mass and bolometric
luminosity, IRAS~22198$+$6336 was classified as an intermediate-mass
star. 
Since this source is
not detected by near-infrared observation, it is considered to be
under formation and deeply buried in dust cloud, equivalent to a
class 0 object in a low-mass star formation
(Hirota et al. 2008, Sanchez-Monge et al. 2010).
An outflow perpendicular to a rotating disk has been
suggested for this source (Palau et al. 2011).

Our aim of this study is to observe the short-term variation and
reveal the origin of the flux variation of the 6.7~GHz
methanol maser. 
The 6.7~GHz methanol maser has not previously been reported
in an intermediate-mass star-forming region and this source is the first example.
Observations and results are presented
in sections 2 and 3, respectively. Section 4 presents
the discussion, and the paper concludes with section 5.

\section{Observations}
We have annually surveyed the 6.7~GHz methanol masers of
approximately 200 sources between 2004 and 2007 repeatedly.
Our aim is to statistically analyze the flux variation of
the 6.7~GHz methanol maser (Fujisawa et al. in prep.).
IRAS~22198$+$6336, identified as a short-term variability source,
was first detected in our observations at 4.4~Jy
($V_\mathrm{LSR}=-8.6$~km~s$^{-1}$) in 2004 (MJD of the observation day was 53226).
This source was below the detection limit (3~Jy) in 2005 (MJD = 53595),
was redetected at 12.1~Jy ($V_\mathrm{LSR}=-18.0$~km~s$^{-1}$)
in 2006 (MJD = 53985), and again fell below the detection limit in 2007 (MJD = 54341).

The methanol maser in IRAS~22198$+$6336 was monitored in 2011 and 2012 by
the Yamaguchi 32-m radio telescope. The 2011 observations
were performed two times every 3~days during an 83-day period
from September 12 (MJD = 55816) to December 3 (MJD = 55898),
yielding 41 observations (excluding missing data).
Three short-time flux variations (hereafter termed flares)
were detected throughout the observation. To study this
variation in detail, 43 daily observations were performed
during 46~days from September 27 (MJD = 56197) to November
11 (MJD = 56242) in 2012. Usual observations were done once per day
as 2011. If a flare was detected, observations
were increased to approximately 20~times per day throughout
the flare duration. During these sub-daily observations,
the maser emission of Cep A, located close to IRAS~22198$+$6336,
was alternately monitored with the target source for flux
calibration. Two sub-daily sessions were performed during
the 2012 observation period. The observation system,
calibration method, and accuracy of the data are reported in
Fujisawa et al. (2012). The observing bandwidth, number of
frequency channels, and velocity resolution were 4~MHz, 4096,
and 0.044~km~s$^{-1}$, respectively. The $1\sigma$ rms noise
level was 1.4~Jy with an integration time of 14~min.
In 2012, the observation bandwidth was 8~MHz, and the
number of frequency channels was 8192. Other parameters
were those used in the 2011 observation.

In addition, three observations were made with
the Hitachi 32-m radio telescope (Yonekura et al. 2013);
December 30, 2012 (MJD = 56291), January 1, 2013 (MJD = 56293),
and January 11, 2013 (MJD = 56303).
The observation bandwidth and number of frequency channels
were 8~MHz and 8192, respectively. The system noise temperature
was 30~K and the $1\sigma$ noise level was 0.3~Jy,
with 5~min integration time.

\section{Results}
\subsection{Results of 2011}
\begin{figure}
  \begin{center}
    \FigureFile(140mm,140mm){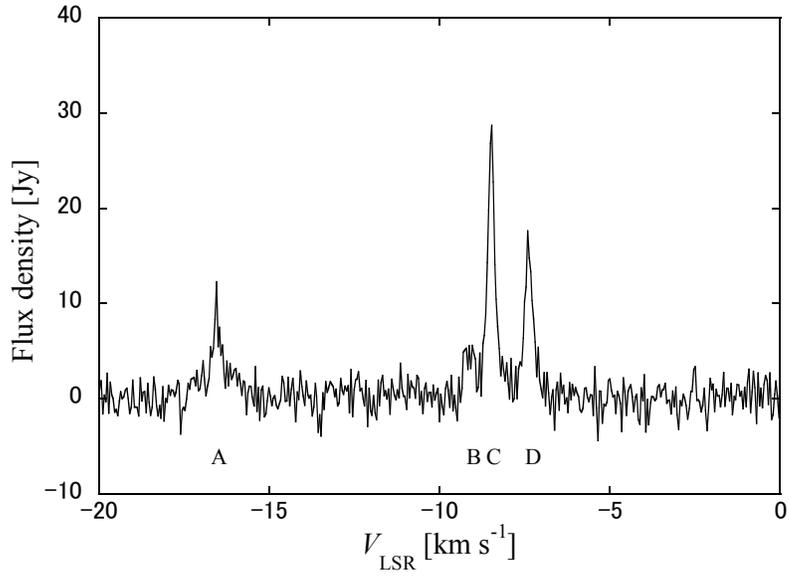}
    \FigureFile(140mm,140mm){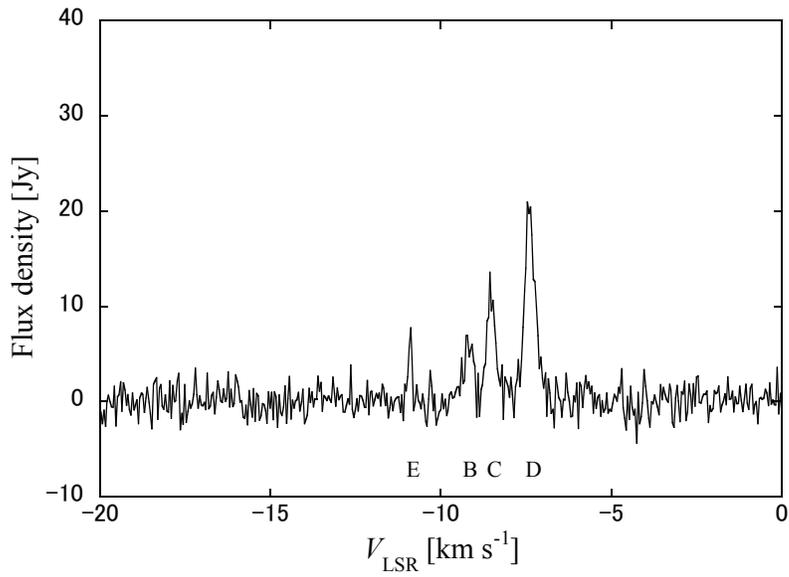}
  \end{center}
  \caption{Upper: Spectrum of the 6.7 GHz methanol maser of IRAS 22198$+$6336
observed on MJD = 55819 during the first flare in 2011.
Lower: Spectrum observed in 2012 (MJD = 56235) during the second flare.}\label{fig:fig1}
\end{figure}

\begin{figure}
  \begin{center}
    \FigureFile(140mm,140mm){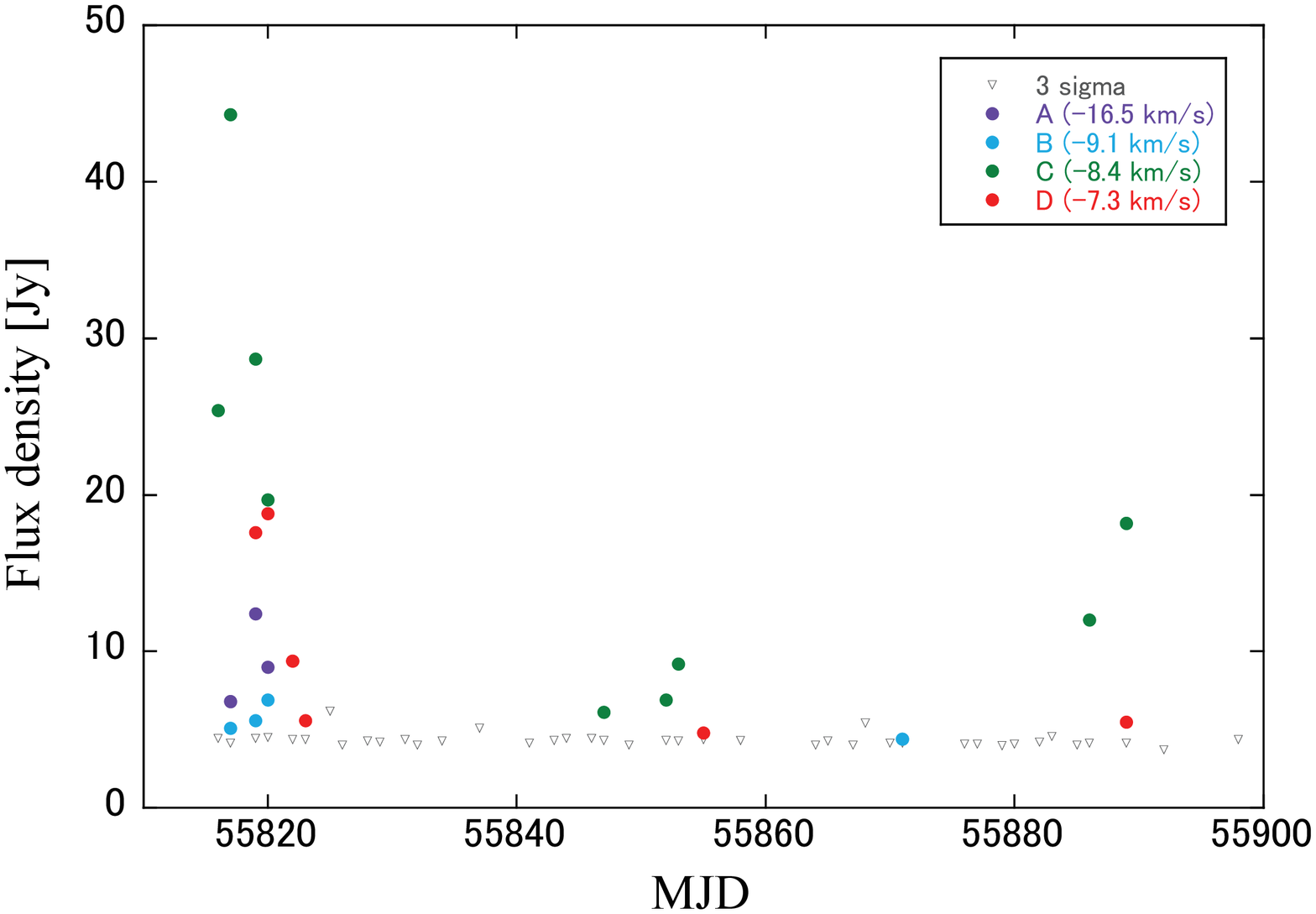}
    \FigureFile(140mm,140mm){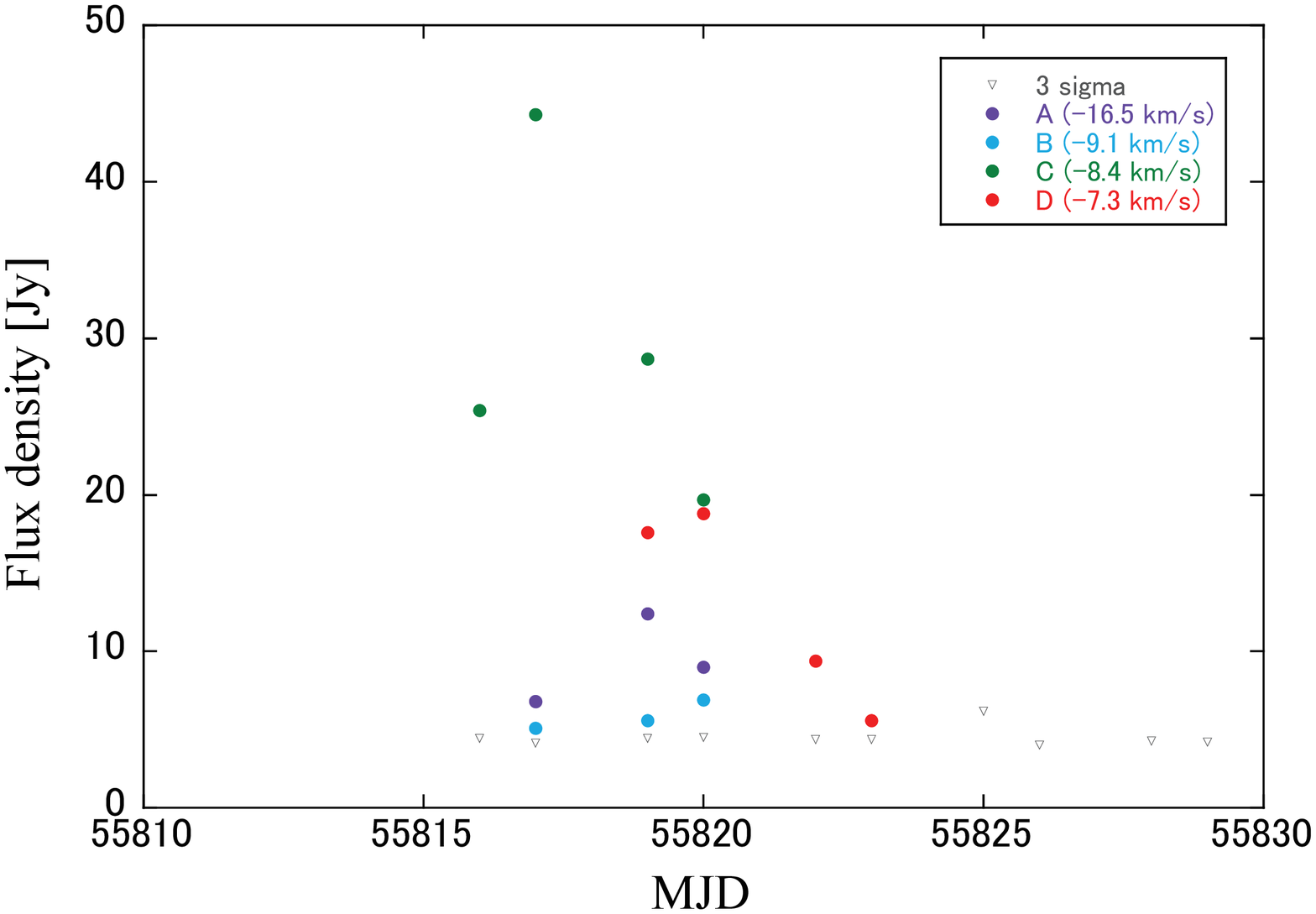}
  \end{center}
  \caption{Upper: Flux variation observed in 2011. Spectral components are distinguished by color: purple = A, cyan = B, green = C, and red = D. Gray triangles represent the 3~$\sigma$ upper limit (approximately 4 Jy).
Lower: A close-up of the first flare.
}\label{fig:fig2}
\end{figure}

Flares were detected in three sub-periods of the observation
period (MJD = 55816$-$55823, 55847$-$55855, and 55886$-$55889).
No maser spectrum with flux density exceeding $3\sigma$ was detected
outside of these flare periods. A typical flare spectrum,
recorded on MJD = 55819, is shown in Figure~\ref{fig:fig1} (upper panel).
Four spectral components, A ($V_\mathrm{LSR}=-16.5$~km~s$^{-1}$),
B ($-$9.1~km~s$^{-1}$), C ($-$8.5~km~s$^{-1}$), and D ($-$7.3~km~s$^{-1}$),
were detected at this time. Figure~\ref{fig:fig2} (upper panel) shows the variation of
the spectral components throughout the 2011 observation.
Here, the horizontal and vertical axes specify the MJD and
flux density, respectively. Although the flux density of
component B is $3.2\sigma$ at MJD = 55871, this signal is disregarded
since it is close to the detection limit and is absent on other days.

The flux variation of all components was intermittent,
and the three flares occurred at equal intervals of
approximately 35~days. During a flare, the four spectral
components varied almost synchronously, although their
velocities were very different (approximately 9.2~km~s$^{-1}$).
A flare lasted for approximately seven~days.
Lower panel of Figure~\ref{fig:fig2} shows a close-up of
the first flare in 2011.
Although the sparse
sampling precluded a detailed light curve, the peak time of each
component was clearly different; only component C is observed on MJD = 55816.
The flux density of component C was its maximum but component D was not detected on the next day (MJD = 55817).
Three days later (MJD = 55820), flux density of component D peaked,
while the flux density of component C was down below the half of its peak.
The preceding variation of C to D was also observed in the second and the third flares.
Peak flux densities also
differed among the flares; during the three flare periods,
the flux density of component C was 44, 10, and 19~Jy.

\subsection{Results of 2012}
During the 2012 observation period, two flares were observed
at MJD = 55198$-$56202 and 56231$-$56238. During the second flare,
IRAS~22198$+$6336 was observed approximately 20~times per day.
Spectral components B, C, and D were detected during
the first flare, while B, C, D, and an additional component E
($-11.0$~km~s$^{-1}$) were detected in the second flare.
Component A was not detected in 2012. A spectrum taken
during the second flare (MJD = 56235) is shown in Figure~\ref{fig:fig1} (lower panel).
The flux variation throughout the 2012 observation
and close-up figures of each flare are shown in
Figure~\ref{fig:fig3}.
Again, component B was detected at $3.1\sigma$
on MJD = 56208, and was disregarded for the reasons mentioned in \S 3.1.

As found in the 2011 observations, two flares separated by 35~days
appeared in 2012. Again, the four spectral components synchronously
flared.
Frequent observation in 2012 revealed the details of the
flux variation. During the first flare (Figure 3 middle panel), component C showed
bursting behavior, increasing rapidly from $<4.4$~Jy ($3\sigma$ upper-limit)
to 17 Jy within one day and
slowly declining over the following four~days.
Component D showed almost symmetrical variation in time,
and the peak time of D was two days later than that of C.
In the second flare, all components 
exhibited almost symmetrical rise and fall of flux density (Figure 3 lower panel).
The peak time clearly differed among the spectral components.
To quantitate these time differences, the light
curves were fitted to Gaussian functions as shown in Figure 3 (lower panel) and their parameters
determined. The derived parameters with their formal errors are presented
in Table~\ref{tab:tab1}. 
The peak time of components B, D, and E
were coincident within 0.5 days, while the peak of C differed from D by 1.8~days.
The variation of C precede that of D
as the same tendency was observed in the other flares in 2011 and 2012.
The average FWHM of the best-fit light curves of C and D was 4~days
that represent a typical time scale of the flare in IRAS~22198+6336.
The relative amplitude in Table~\ref{tab:tab1}, a measure of variability amplitude, is defined as 
\begin{equation}
R = \frac{S_\mathrm{max} - S_\mathrm{min}}{S_\mathrm{min}}
\end{equation}
where $S_\mathrm{max}$ and $S_\mathrm{min}$ are maximum and minimum of the flux density
(van der Walt et al. 2009).
$S_\mathrm{min} = 4.4$ Jy is used here as an upper limit.
The strong variability of IRAS~22198+6336 is remarkable
since the relative amplitude of the flare exceeding 4 occurs in time scale of only 4~days.
Peak flux densities differed among the flares. In the first and
second flares of 2012, component C peaked at 17 and 28~Jy, respectively.

\begin{figure}
  \begin{center}
    \FigureFile(140mm,140mm){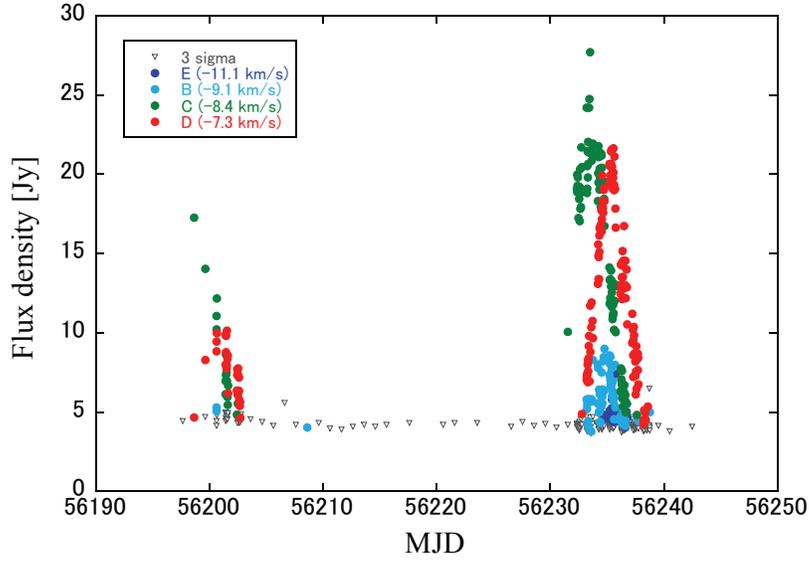}
    \FigureFile(140mm,140mm){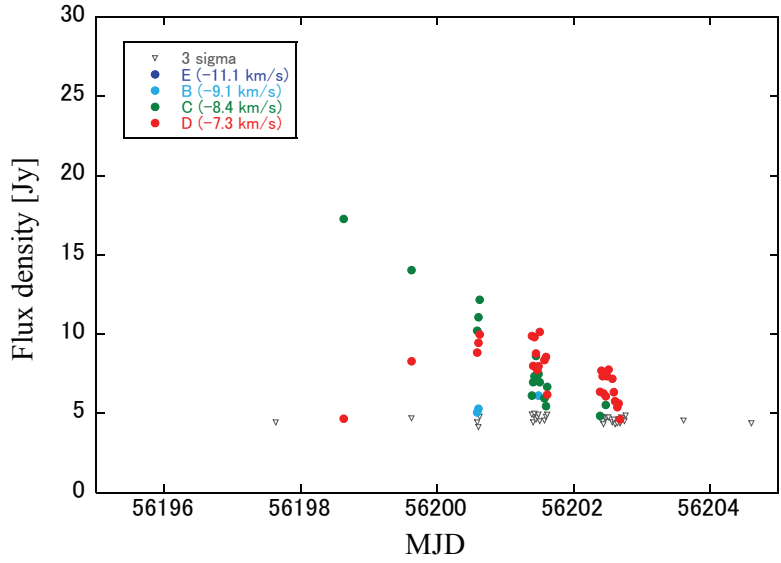}
  \end{center}
  \caption{Upper: flux variation observed in 2012. Spectral components are distinguished by color, as in Figure~\ref{fig:fig2}. Component E is represented by dark blue. Middle: detail of the first flare. Lower:  detail of the second flare with best-fit curves for components C and D.
}\label{fig:fig3}
\end{figure}

\setcounter{figure}{2}

\begin{figure}
  \begin{center}
    \FigureFile(140mm,140mm){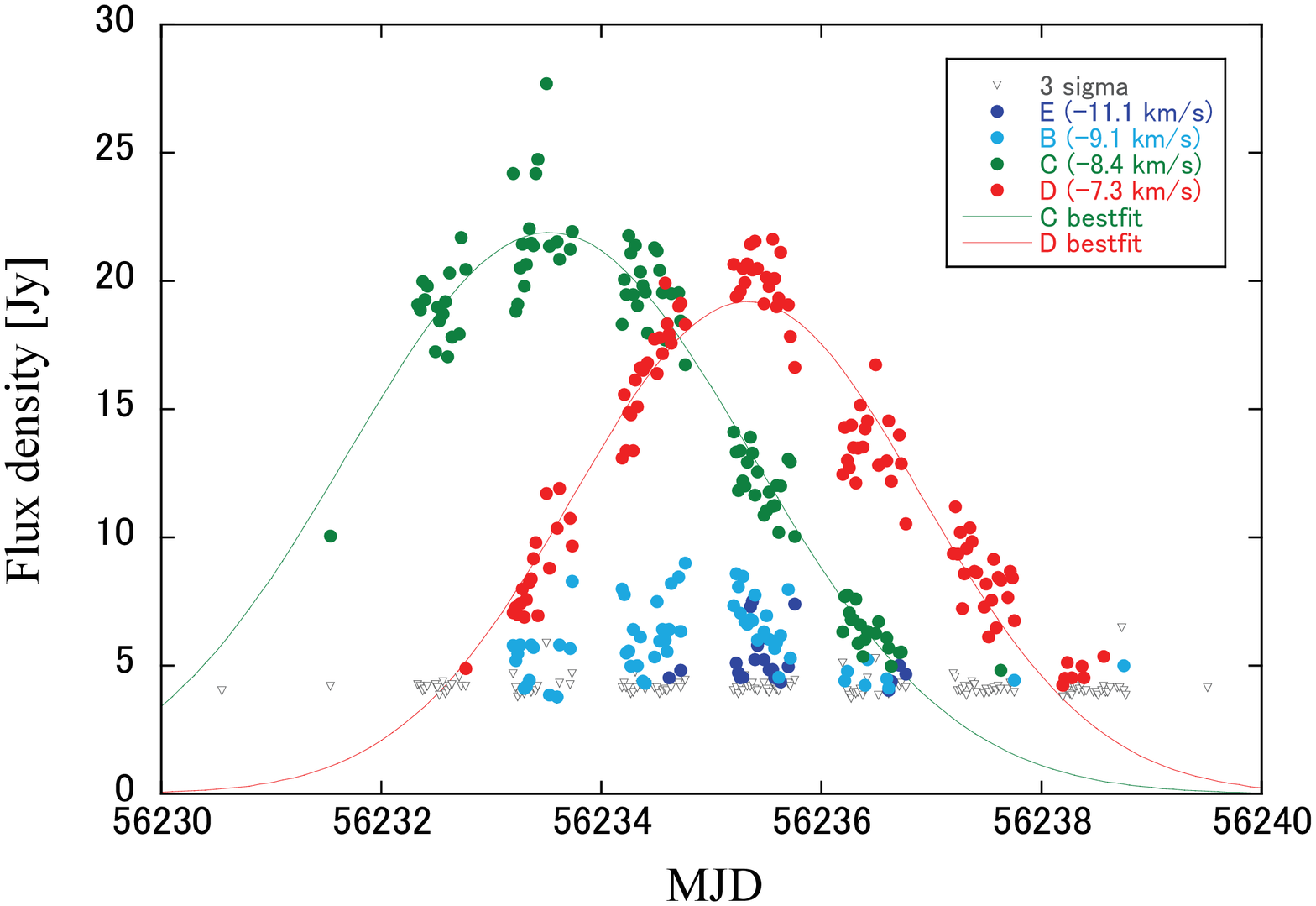}
  \end{center}
  \caption{cont.}
\end{figure}

\begin{table*}
\begin{center}
\caption{Variability parameters of spectral components during the second flare in 2012.}
\label{tab:tab1}
 \begin{tabular}{ccccc}
  \hline\hline
  Component      & Peak flux         & Relative  & Peak time            & Time scale of      \\
                 & density           & amplitude &                      & variability (FWHM) \\
                 & [Jy]              &           & [MJD]                & [days]             \\
  \hline
  B              & \hspace{1.7mm}6.5 & $>0.5$    & 56235.09~$\pm$~0.19  &   6.15~$\pm$0.80   \\
  C              & 21.9              & $>4.0$    & 56233.53~$\pm$~0.04  &   4.31~$\pm$0.09   \\
  D              & 19.2              & $>3.4$    & 56235.33~$\pm$~0.02  &   3.72~$\pm$0.07   \\
  E              & \hspace{1.7mm}5.4 & $>0.2$    & 56235.61~$\pm$~0.29  &   4.29~$\pm$1.48   \\
  \hline
%  \multicolumn{4}{@{}l@{}}{\hbox to 0pt{\parbox{85mm}{\footnotesize
%      Notes. This list is just a sample; it has no meaning.
%    }\hss}}
  \end{tabular}
\end{center}
\end{table*}

\subsection{Observations with Hitachi 32-m radio telescope}
\begin{figure}
  \begin{center}
    \FigureFile(140mm,140mm){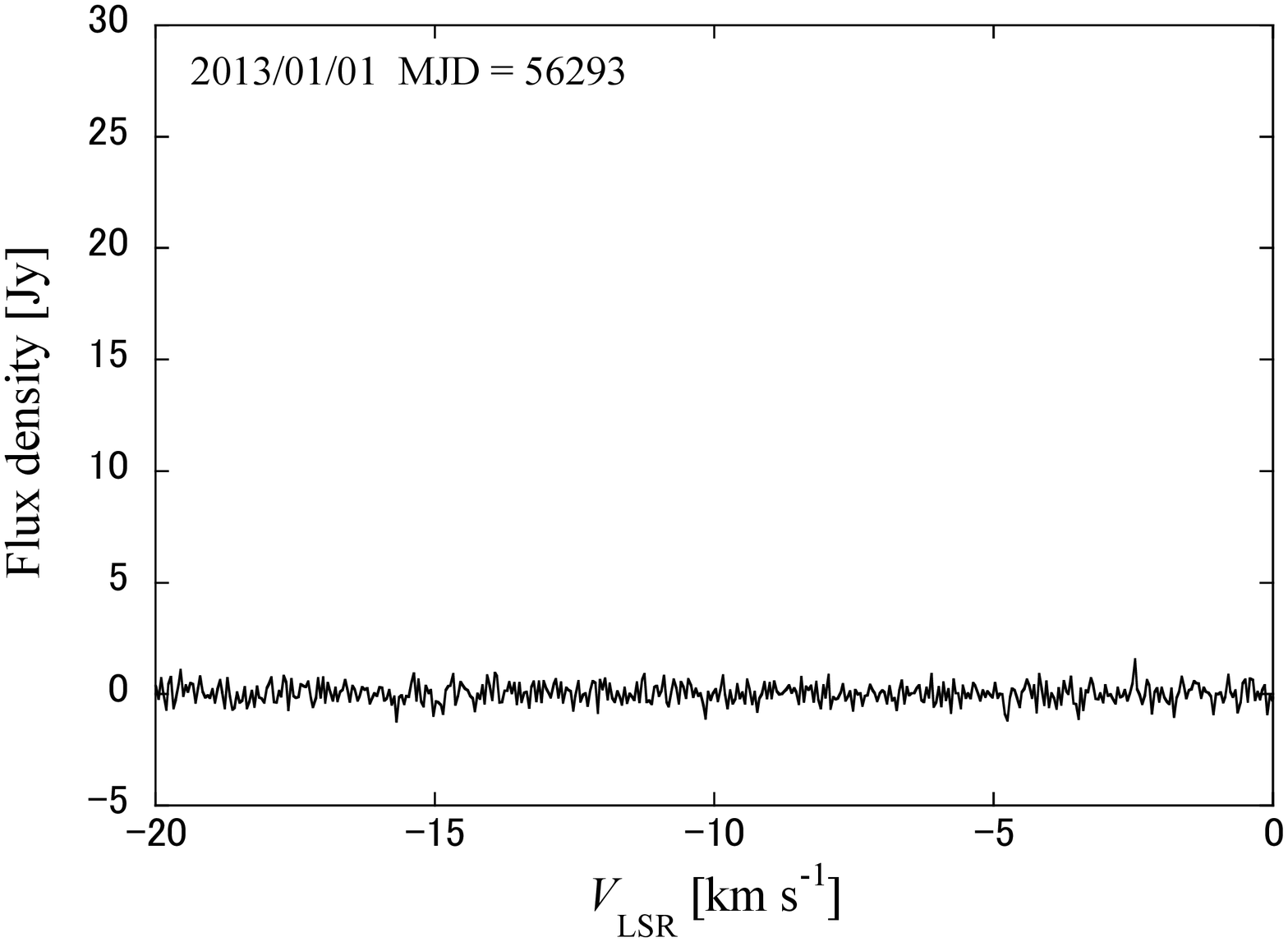}
    \FigureFile(140mm,140mm){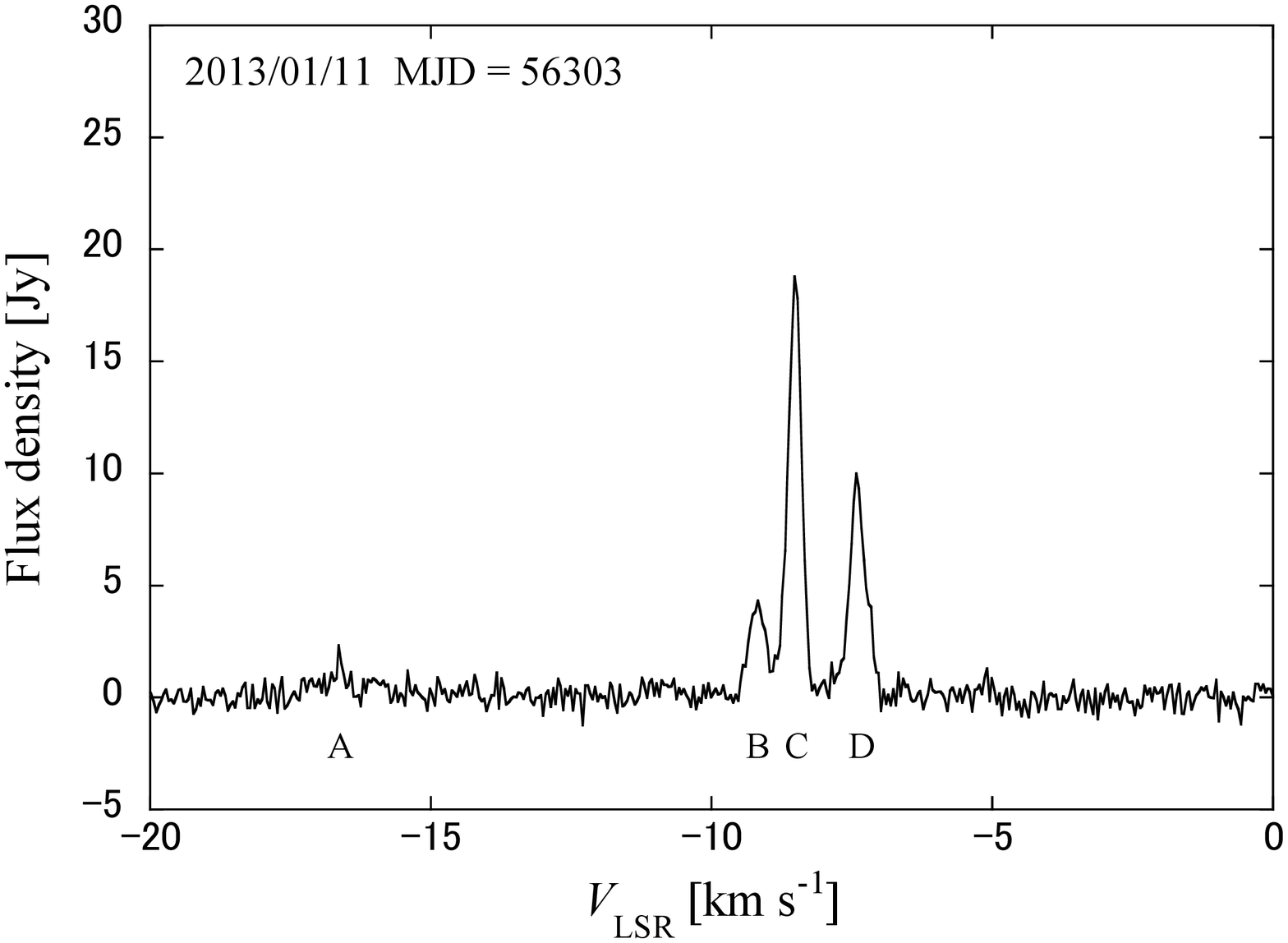}
  \end{center}
  \caption{Spectra of IRAS 22198$+$6336 observed by Hitachi 32-m.
Upper: nondetection (MJD = 56293) and lower: flare (MJD = 56303).
}\label{fig:fig4}
\end{figure}

Three observations were performed by the Hitachi 32-m radio
telescope. No spectral component exceeding the $3\sigma$
upper limit (1.3~Jy) was detected during the first two
observations (MJD = 56291 and 56293). The third observation
at MJD = 56303 detected components A, B, C, and D. Component
E was not detected. Figure~\ref{fig:fig4} shows the spectra at MJD = 56293
and 56303. Component C yielded the highest peak flux density
of 19~Jy.

\subsection{The variation in 2011, 2012, and 2013}
\begin{figure*}
  \begin{center}
    \FigureFile(200mm,200mm){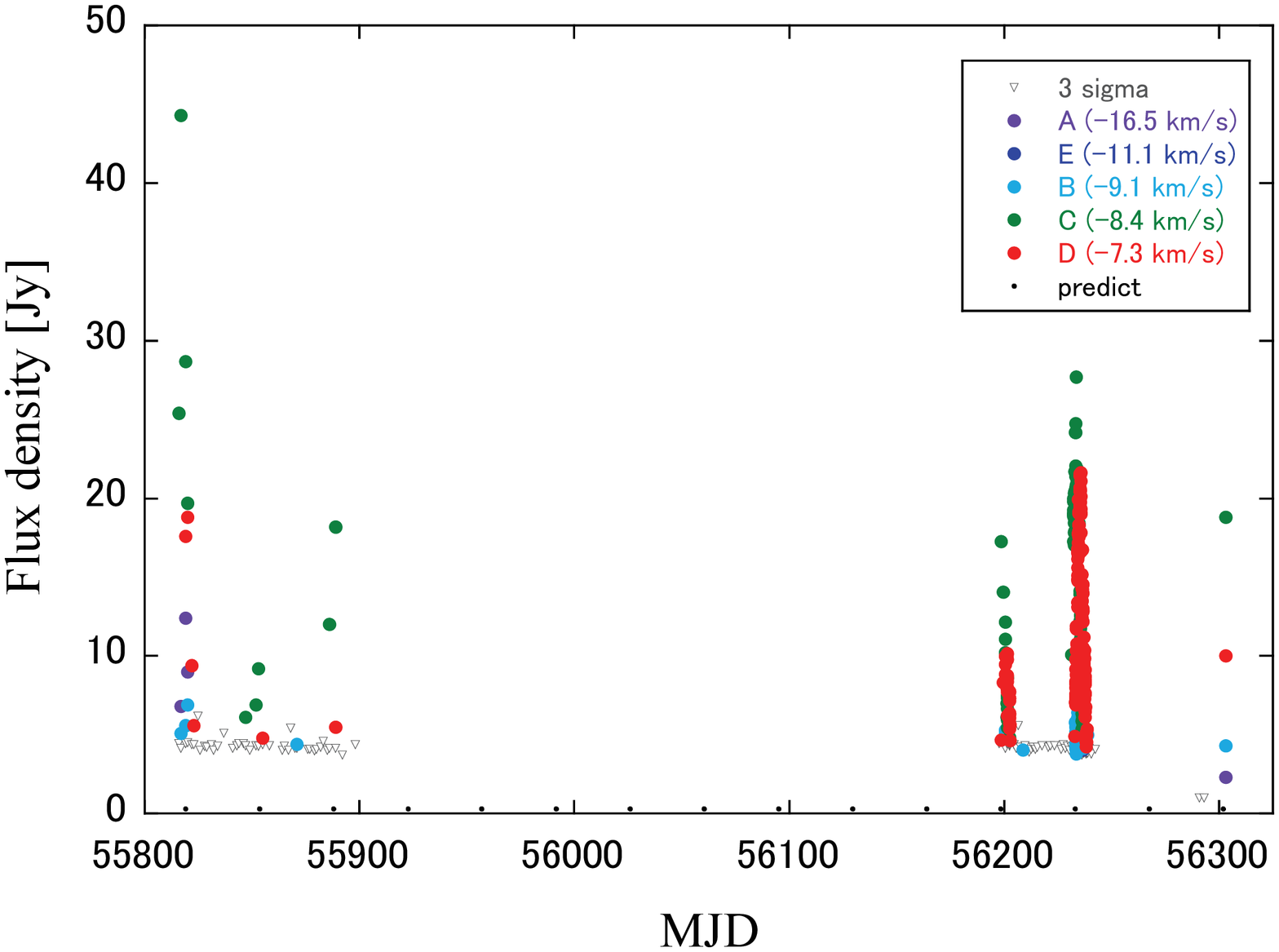}
  \end{center}
  \caption{Flux variation of IRAS 22193$+$6336 observed in 2011, 2012, and 2013.
Evenly spaced dots along the MJD axis indicate dates of expected flares.
}\label{fig:fig5}
\end{figure*}

The flux variations of IRAS~22198$+$6336 observed in 2011, 2012,
and 2013 are collectively shown in Figure~\ref{fig:fig5}. The dots at the bottom
of the figure, indicating expected flare epochs, are spaced at
34.6-day intervals from the first flare observed at MJD = 55818 in 2011. 
The derivation of this interval is discussed in the next section.

\section{Discussions}
\subsection{Periodicity}
It was expected that the flare had arisen with a period of about 35 days
from the observation in 2011 and 2012. 
The day number from the first flare of 2011 (MJD=55818) to the flare of 2013 (MJD=56303) are 485 days.
This is close to 14 times of 35 days. 
The period of 34.6 days is derived by assuming that 485 days is exactly 14 times of the period. 
As shown in Fig. 5, the expected and observed flare epochs well coincide, suggesting
that the flare repeats every 34.6~days.
The dates of the observation performed from 2004 to 2007 were compared
with the date of the flare calculated with the periodicity. 
The calculated MJD is 53224 and the observation MJD is 53226 with only two days difference for the 2004 observation.
For the 2006 observation, the observation MJD of 53985 was just the calculated flare date.
The calculated dates and the observation dates are well coincide for the 2004 and 2006 observations,
and the maser emission was detected at these two observations.
On the other hand, the nearest calculated date is 53605 for the observation date 53595 for the 2005 observation.
For the case of 2007 observation, the nearest date is 54331 for the observation date of 54341.
There are ten days difference between the calculated and observed dates for 2005 and 2007,
and no maser emission was detected for these two observations.
From these facts, it is suspected that the intermittent flare of 6.7~GHz methanol maser of IRAS~22198+6336
with time scale of less than 10 days occur with the period of 34.6 days, 
and this cycle has been kept over eight years or more than 90 cycles from 2004 to 2013. 

This period of 34.6~days is the second
shortest period next to the shortest one 29.5~days for
G12.89$+$0.49 (Goedhart et al. 2009). The third shortest period is
133~days, reported for G338.93$-$0.06 (Goedhart et al. 2007).
Thus, the period of 34.6~days for IRAS~22198$+$6336
is the second example of very short periodic variation sources.
It is notable that all spectral components (A to E) of IRAS~22198+6336 display periodic and synchronized variation,
although other periodic sources tend to exhibit periodicity in limited part of spectrum.

\subsection{Variation pattern}
Both continuous and intermittent variability have been recognized
in the periodic variations of the 6.7~GHz methanol maser.
G12.89$+$0.49, whose period is similar to that of IRAS~22198$+$6336,
displayed a sinusoidal variation pattern. In contrast,
IRAS~22198$+$6336 varied intermittently. According to the Hitachi
32-m observations, the flux density during the quiescent state
is less than or equal to 1.3~Jy.
During the first flare of 2011, the flux density exceeded 40~Jy.
The relative amplitude of the peak flux density to the quiescent flux density exceeds 30.
Similar intermittent
flux variations have been reported for 
G22.356$+$0.066 (Szymczak et al. 2011),
G328.24$-$0.55 (Goedhart et al. 2007),
G37.55$+$0.20 (Araya et al. 2010), and
G9.62$+$0.20E (Goedhart et al. 2003) 
although the relative amplitude of these sources are smaller, i.e., 
5 for G22.356$+$0.066,
1 for G328.24$-$0.55,
10 for G37.55$+$0.20, and
0.2 for G9.62$+$0.20E,
respectively.
Table~\ref{tab:tab2} compares the variability of IRAS~22198$+$6336 with that
of 11 known periodically variable sources. Listed are the period,
variation pattern (intermittent or sinusoidal), variation
range of flux density, and relative amplitude.
The variation patterns are visually judged from the light
curve, and the variation ranges are read from the figures in the
reference papers.
Among the recorded sources, IRAS~22198$+$6336
exhibits the short variation period and the intermittent variation
pattern, with the highest relative amplitude exceeding 30.
Such intermittent variation with the short period uniquely
characterizes the variation of the IRAS~22198$+$6336 maser.

We note that the short-period variation of less than 30 days would be undetectable
if such sources were observed once per month or less frequently.
This means a possibility that there are other short-period sources like IRAS~22198+6336
which are not yet recognized as periodic variable sources. 
Moreover, it is only about 20~\% of the period that IRAS~22198+6336 exhibits the detectable maser emission.
Therefore the detection probability of IRAS~22198+6336 is only 20~\% for single observation.
There might be strong flaring sources like IRAS~22198+6336 not yet discovered by surveys done so far. 
Repeating survey observation would be required for discovery of strong variable sources
similar to IRAS~22198+6336.

\begin{table*}
\begin{center}
\caption{Comparison of variation properties among periodic 6.7 GHz methanol maser sources.}
\label{tab:tab2}
 \begin{tabular}{lccccc}
  \hline\hline
  Name             & Period & Variation pattern  & Variation range & Relative  & Reference \\
                   &  [day] &                    &      [Jy]       & Amplitude & \\
  \hline
  G12.89$+$0.49    & 29.5   & sinusoidal   & 5--20      &  3     & 1          \\
  IRAS22198$+$6336 & 34.6   & intermittent & $<$1.3--44 & $>30$  & this paper \\
  G338.93$-$0.06   & 133    & sinusoidal   & 20--50     &  1.5   & 2          \\
  G22.357$+$0.066  & 179    & intermittent & 1--6       &  5     & 3          \\
  G339.62$-$0.12   & 201    & sinusoidal   & 30--100    &  2     & 2          \\
  G328.24$-$0.55   & 220    & intermittent & 200--400   &  1     & 2          \\
  G37.55$+$0.20    & 237    & intermittent & 0.5--5     &  10    & 4          \\
  G9.62$+$0.20E    & 246    & intermittent & 4500--5500 &  0.2   & 5          \\
  G12.68$-$0.18    & 307    & sinusoidal   &  40--100   &  1.5   & 6          \\
  G188.95$+$0.89   & 404    & sinusoidal   & 500--600   &  0.2   & 2          \\
  G331.13$-$0.24   & 504    & sinusoidal   & 1--20      &  20    & 2          \\
  G196.45$-$1.68   & 668    & sinusoidal   & 20--40     &  1     & 6          \\
  \hline
\multicolumn{5}{@{}l@{}}{\hbox to 0pt{\parbox{120mm}{\footnotesize
     References --- 1: Goedhart et al. (2009), 2: Goedhart et al. (2007), 3: Szymczak et al. (2011), 4: Araya et al. (2010), 5: Goedhart et al. (2003), and 6: Goedhart et al. (2004).
}\hss}}
  \end{tabular}
\end{center}
\end{table*}

The peak times of different spectral components are known to vary
in some periodic variable sources as observed in IRAS~22198$+$6336 (e.g., G12.89+0.49; Goedhart et al. 2009).
This peak time difference could be explained as follows:
the maser spots of each component are distributed
around an exciting source and 
all spectral components intrinsically vary together following the variation of the exciting source.
Then the time lag
would be caused by the different distances between the maser spots
and the observer. Based on this model, Szymczak et al. (2011)
constructed a three-dimensional distribution of maser spots of G22.357+0.066.
In IRAS~22198$+$6336 observations, the peak time of spectral
components C and D differed by 1.8~days, which corresponds to a separation distance of 310~AU
to the line-of-sight.
This distance could be compared with the spatial distribution
of the 6.7~GHz methanol maser emission region of IRAS~22198+6336
derived from the luminosity of the source.
Given that the luminosity
and the exciting dust temperature of the maser region is $450 L_{\solar}$ 
(Hirota et al. 2008) and 100~K (Cragg et al. 2005), respectively,
the distribution of the maser emission region is derived as 330~AU.
This is close to the separation distance of 310 AU derived from the peak time lag.
Future VLBI observation is expected to reveal the spatial distribution of the maser spots of this source.

\subsection{Flaring mechanism}
Several mechanisms responsible for maser variability have been proposed.
The mechanisms proposed are, for example, interstellar
scintillation (Clegg \& Cordes 1991), maser overlap
(Shimoikura et al. 2005), and release of local magnetic energy
(Fujisawa et al. 2012). However, none of these models can explain
the characteristics of the varying 6.7~GHz methanol maser of
IRAS~22198$+$6336, such as its periodicity, synchronization and
time lag of spectral components, and flaring variation pattern.
To explain
these variability characteristics, the model must incorporate a periodic
and intermittent luminosity variation of the exciting source,
which will be reflected in the maser behavior.

According to theory, high-mass protostars that accrete gas
with high accretion rate larger than $10^{-3} M_{\solar}$yr$^{-1}$
should pulsate, thereby periodically altering their radius, temperature,
and luminosity (Inayoshi et al. 2013). Such pulsation should also
yield periodic variation of the excitation state and flux density.
However, the expected variation is sinusoidal, and this model
cannot easily explain the short-term flare. In addition, given
the luminosity ($450 L_{\solar}$) of IRAS~22198$+$6336, the expected period
is approximately one day; alternatively, the luminosity expected
from the period (34.6~days) is $1.4 \times 10^{4} L_{\solar}$. Both quantities
differ from their observed values.

van der Walt et al. (2009) proposed the colliding-wind binary (CWB)
model, which incorporates a binary system with large eccentricity
for interpreting the periodic flare observed in G9.62+0.20E.
When a companion star passes the periastron, the shock induces
ionization photons and heat the surrounding gas and dust.
The observed maser variability would then be caused by
the enhancement of the seed photon 
from the ionized gas. 
The light curve of the periodic flare of G9.62+0.20E showed
a rapid rise and slow decay, and the time scale of decay was about 100 days.
They discussed that the time scale of 100 days can be explained by
the characteristic recombination time of an HII region
which is the origin of the seed photon of the maser emission.
The CWB model
qualitatively explains the observed characteristics of
IRAS~22198$+$6336, such as its periodicity and flaring variability.
Assuming that a companion star of mass $1 M_{\solar}$ orbits the $7 M_{\solar}$ primary
star in 34.6~days, the semi-major axis of the binary is 0.41~AU, which is not
unexceptional in a binary system.

The periodic variability of G9.62+0.20E
were observed at specific spectral components
while the other components did not show the periodicity.
This could be interpreted as that
only the maser clouds locating near side of the HII region
would be strengthened by the seed photon enhancement and show periodic variability. 
Unlike the case of G9.62+0.20E, all the spectral components
with the $9.2$~km~s$^{-1}$ velocity difference flared synchronously in IRAS~22198+6336.
Moreover, there was a time lag of 1.8 days between spectral components.
If the maser emission is enhanced by the seed photons of the HII region
strengthened by the shock induced by the periastron passage,
all the maser emission vary without time lag.
In addition, IRAS~22198+6336 is relatively low luminosity with intermediate-mass exciting star,
there would be no (Ultra-Compact) HII region around the star.

Alternatively, the enhancement of the excitation photon by heated dust
would explain above characteristics
of the periodic flare of IRAS~22198+6336.
The cooling time of optically thick dust radiation is about 1.2 days
(van der Walt et al. 2009) which is shorter than the time scale
of the flare. The dust temperature would
follow the variation of the strength of heating radiation without large delay.
During the first flare of component D and the second flare of C and D (probably B and E too) in 2012,
the light-curve displayed a symmetry in time,
in contrast to that the other sources of intermittent variation
tend to show rapid rise and slow decay
(G9.62$+$0.20E, van der Walt et al. 2009; G22.357$+$0.066,
Szymczak et al. 2011).
Since the interaction of a binary star would arise symmetrically in time to the periastron,
the strength of heating radiation, the dust temperature, and the enhancement
of maser excitation would occur symmetrically in time,
consequently the time symmetry of the light-curve of the flare would be explained.
The flare last only short time and the maser is undetectable at 80\%
of the period suggesting the large eccentricity of a binary star (van der Walt et al. 2009).

\subsection{Water and methanol maser}
Finally we would like to point out the coincidence of the methanol
and water masers in IRAS~22198$+$6336. The 6.7~GHz methanol maser spectrum of
IRAS~22198$+$6336 shows a good correspondence with the 22~GHz water maser
spectrum (Hirota et al. 2008). The spectral components of the water maser
of the almost same velocity exist for the methanol maser components
A ($-$16.5~km~s$^{-1}$), B ($-$9.1~km~s$^{-1}$), and C ($-$8.4~km~s$^{-1}$).
Moreover, the spatial distribution expected from the time lag of
the methanol maser variation is 310~AU, corresponding to 400~mas in the sky,
which is in good agreement with the size of distribution of the water maser
(Hirota et al. 2008). According to these facts, the emission region of
the methanol maser of IRAS~22198$+$6336 may be the same or very close to
the emission region of the water maser.
Bartkiewicz et al. (2011) reported that there are sources
similar to IRAS22198+6336 showing the close velocity distributions
of water and methanol masers.
However, it is revealed by VLBI
observations that water maser shows different spatial
distributions to that of methanol maser in some sources (e.g., Sugiyama et al. 2008, 2011).
The excitation mechanism of the water maser is considered to be collisional
in contrast to the radiation excitation working for the 6.7~GHz methanol maser.
Generally, the water maser has arisen in the region at which outflow of
YSO collides with the surrounding materials.
Although the water maser observed by IRAS~22198$+$6336 shows flux variability,
flare or periodicity like the 6.7~GHz methanol maser have not been reported.
Therefore, coincidence of
the spectrum and the spatial scale may be a coincidence by chance.
Bartkiewicz et al. (2011) noted that water and methanol masers probe different part of the environment of
star forming regions.

Nonetherless, if the distribution of
methanol and water maser is actually spatially close, the cause may
be explained as follows. Once the outflow of YSO collides with the
surrounding materials, the gas is compressed by the shock and its
density and temperature increase, consequently the water maser is excited
and emitted. Simultaneously, the density of the region is high enough
for emitting the methanol maser by radiation excitation by a central star.
This model may also be able to explain that there is a case which methanol
maser appears from the shock region produced by outflow as pointed out by
De Buizer (2003). This model would be verifiable by the spatial distribution
and proper motion observations of methanol and water masers by VLBI monitoring in future.

\section{Conclusions}
We report periodic flaring of the 6.7~GHz methanol maser in an
intermediate-mass star forming region IRAS~22198$+$6336.
IRAS~22198$+$6336 is the first intermediate-mass source detected
by the 6.7~GHz methanol maser. Six flares were detected during
observations undertaken in 2011, 2012, and 2013. The peak flux
density exceeded 40~Jy, but was below the detection limit of
1.3~Jy in the quiescent phase. The flare periodically erupted
every 34.6~days. The time scale of a single flare was as short
as 4~days, and rapid flux variation was observed. Including
IRAS~22198$+$6336, periodic variability of the 6.7~GHz methanol
maser has been reported in 12 sources to date. The periodic
variability of IRAS~22198$+$6336 is uniquely characterized by
its short period and its flaring nature, and is reasonably
explained by the CWB model.

\bigskip

The authors thank Mr. Hiramoto for his assistance to the data analysis,
National Astronomical Observatory of Japan, KDDI Corporation for supporting 
the Yamaguchi 32-m radio telescope.
This work was financially supported in part by Grant-in-Aid for
Scientific Research (KAKENHI) from the Japan Society for
the Promotion of Science (JSPS), No. 24340034.

%%%
% See the manual for the detail.
%%%


\begin{thebibliography}{}
\bibitem[Araya et al.(2010)]{2010ApJ...717L.133A} 
Araya, E.~D., Hofner, P., Goss, W.~M., Kurtz, S., Richards, A.~M.~S., Linz, H., Olmi, L., \& Sewi{\l}o, M.\ 2010, \apjl, 717, L133 
\bibitem[Bartkiewicz et al.(2009)]{2009A&A...502..155B}
Bartkiewicz, A., Szymczak, M., van Langevelde, H.~J., Richards, A.~M.~S., \& Pihlstr{\"o}m, Y.~M.\ 2009, \aap, 502, 155
\bibitem[Breen et~al.(2013)]{2013MNRAS.435..524B}
Breen, S.~L., Ellingsen, S.~P., Contreras, Y., Green, J.~A., Caswell, J.~L., Stevens, J.~B., Dawson, J.~R., \& Voronkov, M.~A.\ 2013, \mnras, 435, 524
\bibitem[Caswell et al.(1995)]{1995MNRAS.272...96C}
Caswell, J.~L., Vaile, R.~A., Ellingsen, S.~P., Whiteoak, J.~B., \& Norris, R.~P.\ 1995, \mnras, 272, 96
\bibitem[Clegg \& Cordes(1991)]{1991ApJ...374..150C} 
Clegg, A.~W., \& Cordes, J.~M.\ 1991, \apj, 374, 150 
\bibitem[Cragg et al.(2005)]{2005MNRAS.360..533C} 
Cragg, D.~M., Sobolev, A.~M., \& Godfrey, P.~D.\ 2005, \mnras, 360, 533 
\bibitem[De Buizer (2003)]{2003MNRAS.341..277D}
De Buizer, J.~M.\ 2003, \mnras, 341, 277
\bibitem[Fujisawa et~al.(2012)]{2012PASJ...64...17F}
Fujisawa, K., Sugiyama, K., Aoki, N., Hirota, T., Mochizuki, N., Doi, A., Honma, M., Kobayashi, H., Kawaguchi, N., Ogawa, H., Omodaka, T., \& Yonekura, Y.\ 2012, \pasj, 64, 17
\bibitem[Goedhart et al.(2003)]{2003MNRAS.339L..33G} 
Goedhart, S., Gaylard, M.~J., \& van der Walt, D.~J.\ 2003, \mnras, 339, 33 
\bibitem[Goedhart et al.(2004)]{2004MNRAS.355..553G} 
Goedhart, S., Gaylard, M.~J., \& van der Walt, D.~J.\ 2004, \mnras, 355, 553 
\bibitem[Goedhart et al.(2007)]{2007IAUS..242...97G}
Goedhart, S., Gaylard, M.~J., \& van der Walt, D.~J.\ 2007, Astrophysical Masers and their Environments, Proceedings of the International Astronomical Union, IAU Symposium, Volume 242, p. 97-101
\bibitem[Goedhart et al.(2009)]{2009MNRAS.398..995G}
Goedhart, S., Langa, M.~C., Gaylard, M.~J., \& van der Walt, D.~J.\ 2009, \mnras, 398, 995 
\bibitem[Hirota et al.(2008)]{2008PASJ...60..961H}
Hirota, T., Ando, K., Bushimata, T., Choi, Y.~K., Honma, M., Imai, H., Iwadate, K., Jike, T., Kameno, S., Kameya, O. et al.\ 2008, \pasj, 60, 961
\bibitem[Inayoshi et al.(2013)]{2013ApJ...769L..20I} 
Inayoshi, K., Sugiyama, K., Hosokawa, T., Motogi, K., \& Tanaka, K.~E.~I.\ 2013, \apj, 769, 20
\bibitem[Menten (1991)]{1991ApJ...380L..75M}
Menten, K.~M. 1991, \apj, 380, 75
\bibitem[Minier et~al.(2000)]{2000A&A...362.1093M}
Minier, V., Booth, R.~S., \& Conway, J.~E.\ 2000, \aap, 362, 1093
\bibitem[Minier et al.(2003)]{2003A&A...403.1095M}
Minier, V., Ellingsen, S.~P., Norris, R.~P., \& Booth, R.~S.\ 2003, \aap, 403, 1095
\bibitem[Palau et~al.(2011)]{2011ApJ...743L..32P}
Palau, A., Fuente, A., Girart, J.~M., Fontani, F., Boissier, J., Pietu, V., Sanchez-Monge, A., Busquet, G., Estalella, R., Zapata, L.~A., Zhang, Q., Neri, R., Ho, P.~T.~P., Alonso-Albi, T., \& Audard, M.\ 2011, \apj, 743, 32 
\bibitem[Sanchez-Monge et al.(2008)]{2008A&A...485..497S} 
Sanchez-Monge, A., Palau, A., Estalella, R., Beltran, M.~T., \& Girart, J.~M.\ 2008, \aap, 485, 497
\bibitem[Sanchez-Monge et al.(2010)]{2010ApJ...721L.107S} 
Sanchez-Monge, A., Palau, A., Estalella, R., Kurtz, S., Zhang, Q., Di Francesco, J., \& Shepherd, D.\ 2010, \apj, 721, 107
\bibitem[Shimoikura et al.(2005)]{2005ApJ...634..459S} 
Shimoikura, T., Kobayashi, H., Omodaka, T., Diamond, P.~J., Matveyenko, L.~I., \& Fujisawa, K.\ 2005, \apj, 634, 459 
\bibitem[Sugiyama et al.(2008)]{2008PASJ...60.1001S}
Sugiyama, K., Fujisawa, K., Doi, A., Honma, M., Isono, Y., Kobayashi, H., Mochizuki, N., \& Murata, Y.\ 2008, \pasj, 60, 1001
\bibitem[Sugiyama et al.(2011)]{2011PASJ...63...53S}
Sugiyama, K., Fujisawa, K., Doi, A., Honma, M., Isono, Y., Kobayashi, H., Mochizuki, N., Murata, Y., Sawada-Satoh, S., \& Wajima, K.\ 2011, \pasj, 63, 53
\bibitem[Sugiyama et al.(2014)]{2014A&A...562A..82S}
Sugiyama, K., Fujisawa, K., Doi, A., Honma, M., Kobayashi, H., Murata, Y., Motogi, K., Niinuma, K., Ogawa, H., Wajima, K., Sawada-Satoh, S., \& Ellingsen, S. P.\ 2014, \aap, 562, 82
\bibitem[Szymczak et al.(2011)]{2011A&A...531L...3S}
Szymczak, M., Wolak, P., Bartkiewicz, A., \& van Langevelde, H.~J.\ 2011, \aap, 531, L3
\bibitem[Tafalla et al. (1993)]{1993ApJ...403..175T}
Tafalla, M., Bachiller, R., \& Martin-Pintado, J.\ 1993, \apj, 403, 175
\bibitem[van der Walt et al.(2009)]{2009MNRAS.398..961V} 
van der Walt, D.~J., Goedhart, S., \& Gaylard, M.~J.\ 2009, \mnras, 398, 961 
\bibitem[Xu et al.(2008)]{2008A&A...485..729X}
Xu, Y., Li, J.~J., Hachisuka, K., et al.\ 2008, \aap, 485, 729
\bibitem[Yonekura et al.(1997)] {1997ApJS..110...21Y}
Yonekura, Y., Dobashi, K., Mizuno, A., Ogawa, H., \& Fukui, Y.\ 1997, \apjs, 110, 21
\bibitem[Yonekura et al.(2013)] {2013ASPC..476..415Y}
Yonekura, Y., Saito, Y., Saito, T., et al. 2013,
ASP Conf. Ser. 476: New Trends in Radio Astronomy in the ALMA Era:
The 30th Anniversary of Nobeyama Radio Observatory (eds. R. Kawabe, N. Kuno, S. Yamamoto), 415

\end{thebibliography}
\end{document}